\documentclass[nonacm,sigconf,screen]{acmart}

\usepackage[linesnumbered,ruled,vlined]{algorithm2e}
\usepackage{booktabs}
\usepackage[nameinlink]{cleveref}  
\usepackage{gensymb}
\usepackage{listings}
\usepackage[binary-units]{siunitx}
\usepackage{todonotes}

\DeclareSIUnit\cycles{cycles}
\setcopyright{none}
\copyrightyear{2021}
\acmYear{2021}
\acmDOI{10.1145/1122445.1122456}

\acmConference[Woodstock '18]{Woodstock '18: ACM Symposium on Neural
  Gaze Detection}{June 03--05, 2018}{Woodstock, NY}
\acmBooktitle{Woodstock '18: ACM Symposium on Neural Gaze Detection,
  June 03--05, 2018, Woodstock, NY}
\acmPrice{15.00}
\acmISBN{978-1-4503-XXXX-X/18/06}

\begin{document}

\title[Towards Retina-Quality VR Video Streaming: \SI{15}{\ms} Could Save You 80\% of Your Bandwidth]{Towards Retina-Quality VR Video Streaming: \\15\,ms Could Save You 80\% of Your Bandwidth}



\author{Luke Hsiao, Brooke Krajancich, Philip Levis, Gordon Wetzstein, and Keith Winstein}
\affiliation{%
    \institution{Stanford University}
    \streetaddress{Electrical Engineering Department, Stanford University}
    \city{Stanford}
    \state{California}
    \country{USA}
    \postcode{94305}
}
\email{{lwhsiao@cs.,brookek@,pal@cs.,gordon.wetzstein@,keithw@cs.}stanford.edu}

\renewcommand{\shortauthors}{Hsiao, et al.}

\begin{abstract}
    Virtual reality systems today cannot yet stream immersive, retina-quality virtual reality video over a network.
    One of the greatest challenges to this goal is the sheer data rates required to transmit retina-quality video frames at high resolutions and frame rates.
    Recent work has leveraged the decay of visual acuity in human perception in novel gaze-contingent video compression techniques.
    In this paper, we show that reducing the motion-to-photon latency of a system itself is a key method for improving the compression ratio of gaze-contingent compression.
    Our key finding is that a client and streaming server system with sub-\SI{15}{\ms} latency can achieve \SI{5}{\times} better compression than traditional techniques while also using simpler software algorithms than previous work.
\end{abstract}



\begin{teaserfigure}
    \renewcommand{\thefootnote}{\scriptsize\arabic{footnote}}
    \centering
    \includegraphics[width=0.95\textwidth]{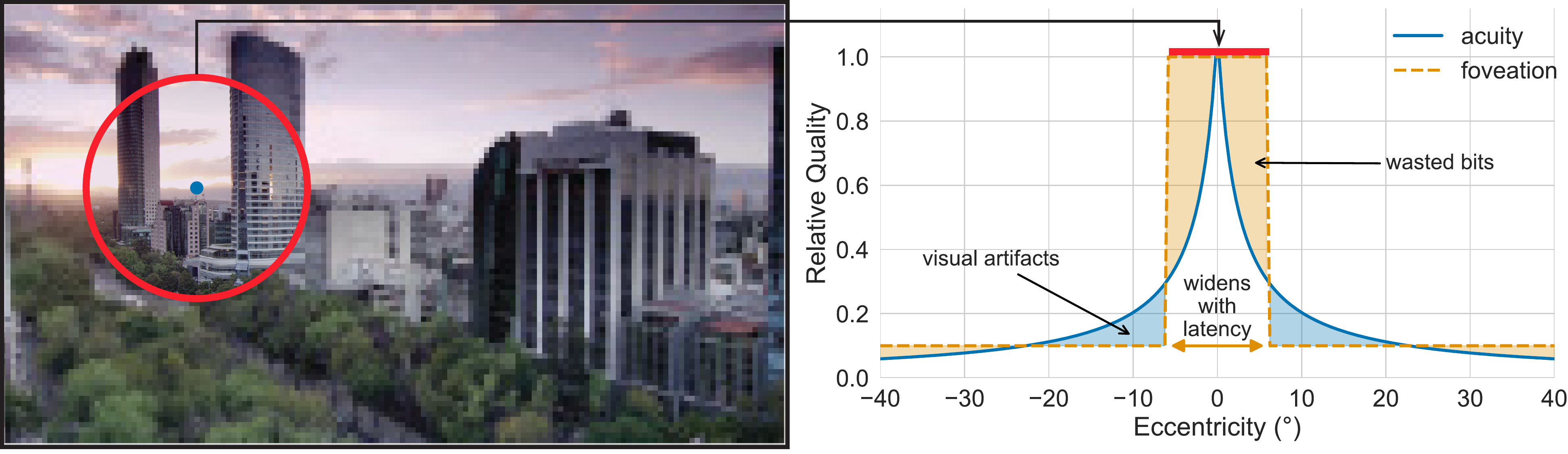}
    \caption{%
        One reason virtual reality systems today cannot yet deliver retina-quality video experiences is due to bandwidth limitations.
        To reduce data rates, recent work uses the decay of visual acuity in human perception for foveated video compression, keeping a small region of high resolution while decaying quality in the periphery (left)\protect\footnotemark.
        We show that decreasing motion-to-photon latency benefits foveated video compression and enables minimally-sized regions of high resolution (right).
    }\label{fig:teaser}
\end{teaserfigure}

\maketitle

\footnotetext[1]{Note that the quality reduction has been exaggerated for illustration purposes.}
\addtocounter{footnote}{1}

\section{Introduction}%
\label{sec:intro}

Virtual reality (VR) video strives to offer immersive experiences through high fidelity, \ang{360} display of recorded content.
Doing so requires streaming video at both high resolutions and frame rates across large fields of view with constrained computational power and bandwidth.
Today's VR systems, unable to achieve this, stream videos below retina resolution, at low frame rates, or both.

Several practical challenges stand in the way of achieving immersive, retina-quality VR video.
First, only powerful GPUs are capable of decoding high resolution video frames at, or greater than, the \SI{90}{\Hz} or higher refresh rates that are essential for VR~\cite{cuervo2018creating}.
Second, modern VR display hardware does not yet support retina-quality pixel densities.
Consumer headsets today only reach about \SI{20}{\percent} of that goal\footnote{For example, one pixel of the HTC VIVE Pro is approximately \ang{;4;35} of visual angle, or \SI{\sim{} 5}{\times} larger than the minimum angle of resolution in the foveola}.
Third, the poor performance of existing systems is due in part to the sheer amount of data our retinal acuity requires.
In this paper, we focus on this third challenge.

Consider an \emph{uncompressed} 5.7K (\SI{5760 x 2880}{px}), \SI{360 x 180}{\degree}, VR video---the highest resolution supported by \ang{360} cameras today.
Setting aside the immense bandwidth requirements of streaming uncompressed 5.7K video (\SI{\sim{}7}{\giga\bit\per\second}), this would still only achieve \SI{16}{samples\per\degree}, just \SI{27}{\percent} of the \SI{60}{samples\per\degree} standard for retina quality.
Since streaming services like YouTube encode 5.7K video at \SIrange{15}{30}{\mega\bit\per\second} (\SI{>230}{\times} smaller than the uncompressed bitrate), the resolution after compression is even worse.
Achieving retina-quality VR video with traditional techniques would require a huge increase in bitrate (higher resolutions and less compression); the bandwidth requirements alone are a barrier.

This challenge has inspired perceptually-motivated graphics; a complementary field of work that exploits the limitations of human perception to reduce bandwidth or computation.
In particular, these techniques use the fact that our visual acuity (or ability to resolve spatial detail) is highest in the region of the retina called the fovea and drops quickly with eccentricity (or distance from the fovea).
Combined with eye tracking, this knowledge is used to degrade rendering quality~\cite{guenter_foveated_2012, patney2016towards, friston2019perceptual}, level-of-detail~\cite{luebke2001perceptually, murphy2001gaze, ohshima1996gaze}, or display resolution~\cite{kim2019foveated, tan2018foveated} in regions that fall on a user's periphery, thus reducing bandwidth without perceivable quality degradation.
For streaming \ang{360} video, related work also utilizes techniques such as adapting encoding parameters~\cite{guan2019pano}, predicting a user's field of view~\cite{sun2020flocking} and upscaling highly compressed video using super-resolution~\cite{chen2020super} (\Cref{fig:teaser} shows an example).
We focus on foveated video compression, which seeks to compress a sequence of frames while modeling visual acuity decay to concentrate the allocation of bits in the encoded video to the foveal region~\cite{romero2018foveated,kaplanyan2019deepfovea,illahi2020cloud}.

While the compression benefits of foveated techniques are significant, they are fundamentally limited by the \emph{motion-to-photon latency} of the system.
This latency is the time between a change in the viewer's gaze and the resulting change in the display's pixels.
Larger latencies introduce larger uncertainty about the viewer's gaze position and consequently require a larger foveal region to avoid perception of the degradation applied in the periphery.

We present a study on the relationship between a system's motion-to-photon latency and the bitrate required to display a gaze-contingent video without degrading its perceived quality.
We use a desktop setup as a proxy for future high-frame-rate, low-latency, retina-resolution VR systems.
Our key finding is that with sub-\SI{15}{\ms} latency, we improve on the bitrate of traditional compression techniques by \SI{5}{\times} while using simpler software techniques than previous work.
We also find that some compression gains are only achieved with latencies under \SI{45}{\ms}; above this threshold, the benefits of reducing latency are less pronounced.
We believe using gaze-contingent compression with low-latency systems is a key step towards realizing truly immersive VR experiences.

\paragraph{Contributions} This paper makes these contributions:
\begin{itemize}
    \item We build a video streaming system using foveated video compression.
        The display reacts to gaze changes within \SI{15}{\ms}, over \SI{3}{\times} lower than previously demonstrated in VR HMDs.
    \item Through a user study using our low-latency prototype, we derive perceptual insights about the relationship between system latency and the bitrate required to display a foveated video without noticeable quality degradation.
    \item We find that low latency can reduce the number of video bits a system must receive and decode by \SI{2}{\times}, but only when latency is far below previously proposed thresholds.
\end{itemize}

We directly focus on the impact of motion-to-photon latency.
As a result, our design has a few important limitations.
First, our prototype system does not include the latency introduced by separating the client and server with a realistic network.
The need for low server-to-client latencies means that a video encoder would need to be located near the client at the network edge; this might be a use case for edge computing.
Second, our prototype uses an encoder-in-the-loop approach to perform video compression and streaming in real-time.
This approach has a higher computational cost than those that pre-encode chunks of video and requires the server to encode video for each individual viewer.
Last, we evaluate our system using an eye tracker and display that are among the fastest available today;
comparable performance is unavailable on the consumer market or in current head-mounted displays.

\section{Background and Related Work}%
\label{sec:background}

\subsection{Human Perception}%
\label{sec:perception}

The human visual system has a field of view of approximately \ang{220} horizontally by \ang{135} vertically~\cite{knapp1938introduction}.
Yet only a small region (\SI{\sim~1.5}{\degree}), called the fovea, is capable of resolving spatial detail as fine as \SI{60}{\cycles\per\degree}~\cite{deering1998limits}.
Outside the fovea, the distribution of retinal components and refractive lens effects change rapidly, resulting in decreased visual acuity~\cite{thibos1987retinal}, less sensitivity to color~\cite{anderson1991human, hansen2009color}, and limited stereoscopic depth discrimination~\cite{siderov1995stereopsis}, as well as increased sensitivity to flicker~\cite{krajancich2021perceptual, hartmann1979peripheral} in our peripheral visual field.

The eyes make short, rapid movements called saccades to scan visual scenes with the high-resolution fovea.
While these ballistic-like movements can occur at speeds of up to \SI{\sim~900}{\degree\per\second}~\cite{carpenter1988movements}, the temporary suspension in perception (referred to as saccadic suppression) that occurs a short period before, during, and after the eye movement (totaling \SIrange{50}{200}{\ms}~\cite{ross2001changes}) reduces the challenge they pose to gaze-contingent systems.
However, even during fixation the eyes involuntarily move, albeit slower (\SI{\sim~50}{\arcminute\per\second}~\cite{rucci2015control}), exploring fine detail with a random-walk-like pattern referred to as ocular drift and correcting the fixation position with microsaccades.
During fixation, there is also a high frequency component referred to as ocular tremor (see~\cite{kowler2011eye} for a detailed review).

\subsection{Foveated Video Compression}%
\label{sec:foveated_compression}

This knowledge of the human visual system, coupled with real-time eye tracking, has given rise to foveated graphics techniques that imperceptibly degrade the peripheral image to improve efficiency (i.e., reducing bandwidth or computation).
For example, foveated graphics improves efficiency by reducing the number of vertices or fragments a GPU has to sample, ray trace, shade, or transmit to the display~\cite{koulieris2019near}.
The most prominent approach is perhaps foveated rendering~\cite{guenter_foveated_2012, patney2016towards, friston2019perceptual} and display~\cite{kim2019foveated, tan2018foveated}, where images and videos are rendered, transmitted, or displayed with spatially varying resolutions without affecting the perceived image quality.
Related approaches also use gaze location to vary bit-depth~\cite{mccarthy2004sharp}, shading or level-of-detail~\cite{luebke2001perceptually, murphy2001gaze, ohshima1996gaze}, or reconstruct content from sparse samples~\cite{kaplanyan2019deepfovea} outside of the foveal region.

These ideas have also been applied to video compression.
Traditional video compression removes temporal and spatial redundancy in a sequence of video frames.
Foveated video compression builds on these techniques by using real-time gaze information to concentrate data allocation in an encoded video to the foveal region, achieving better compression in the periphery.

There are many approaches for foveated video compression.
Lee~et~al.~\cite{sanghoon_lee_foveated_2001} use a nonuniform filtering scheme to increase compression.
Specifically, their algorithm maximizes a foveated signal-to-noise ratio (FSNR) using a Lagrange multiplier along curvilinear coordinates.
Illahi~et~al.~\cite{illahi2020cloud} use a similar but simpler approach of varying quantization parameters, compressing peripheral regions more than foveal regions.
Instead of compressing a single video stream, Romero~et~al.~\cite{romero2018foveated} store a video in two resolutions, low and high.
A client first fetches the low-resolution stream, and then streams only the cropped, high-resolution segments based on a viewer's current gaze.
Similarly, Jeppsson~et~al.~\cite{jeppsson2018efficient} divide a video into many small blocks and pre-encodes each block in many different resolutions.
Then, when streaming, the resolutions are chosen on the server based on gaze data and stitched together at the client into three levels of resolution.
Foveated video compression can achieve bitrates that are \SIrange{25}{60}{\percent} of the bitrates of traditional compression algorithms with similar visual quality.

\subsection{Latency}%
Being gaze-contingent, foveated compressions systems are very sensitive to motion-to-photon latency---the time between the eyes moving and the pixels of the display updating with the frame corresponding to the new gaze location.
Yet none of the foveated video compression works described in \Cref{sec:foveated_compression} discuss the impact of latency on their results.

The importance of system latency has been given more attention in foveated rendering, with a number of works measuring the maximal tolerable system latency to be between \SIrange{42}{91}{\ms}, depending on the size of the full resolution foveal image that follows the gaze, the degree of degradation applied to the image, and the type of degradation method used~\cite{guenter_foveated_2012,thunstrom2014passive, stengel_adaptive_2016, albert_latency_2017}.
Similarly, Loschky~et~al.~\cite{loschky2007late} also observed that detection of image artifacts due to foveation in gaze-contingent, multiresolution displays did not change if latency was kept under \SI{60}{\ms}.
However, to the best of our knowledge we are the first to show the significant compression benefits of squeezing system latency below these thresholds in reducing the bitrate needed to produce the same visual quality.

\section{Latency vs.\ Compression}
\label{sec:latency}

\begin{figure}[t]
    \centering
    \includegraphics[width=1.0\columnwidth]{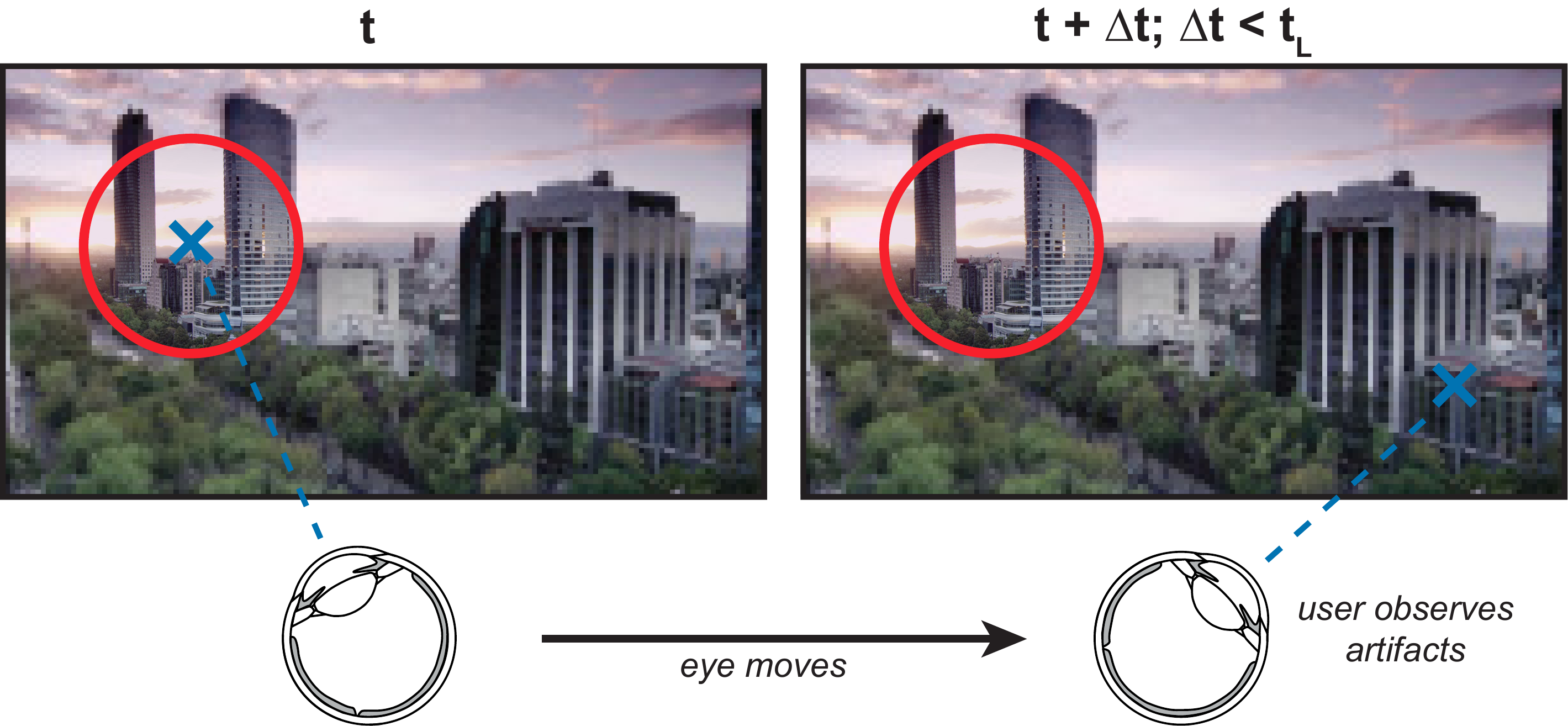}
    \caption{
        A system must compensate for latency by enlarging the foveal region to avoid a viewer's gaze escaping the region before the system can react.
    }\label{fig:approximations}
\end{figure}

\begin{figure}[t]
    \centering
    \includegraphics[width=1.0\columnwidth]{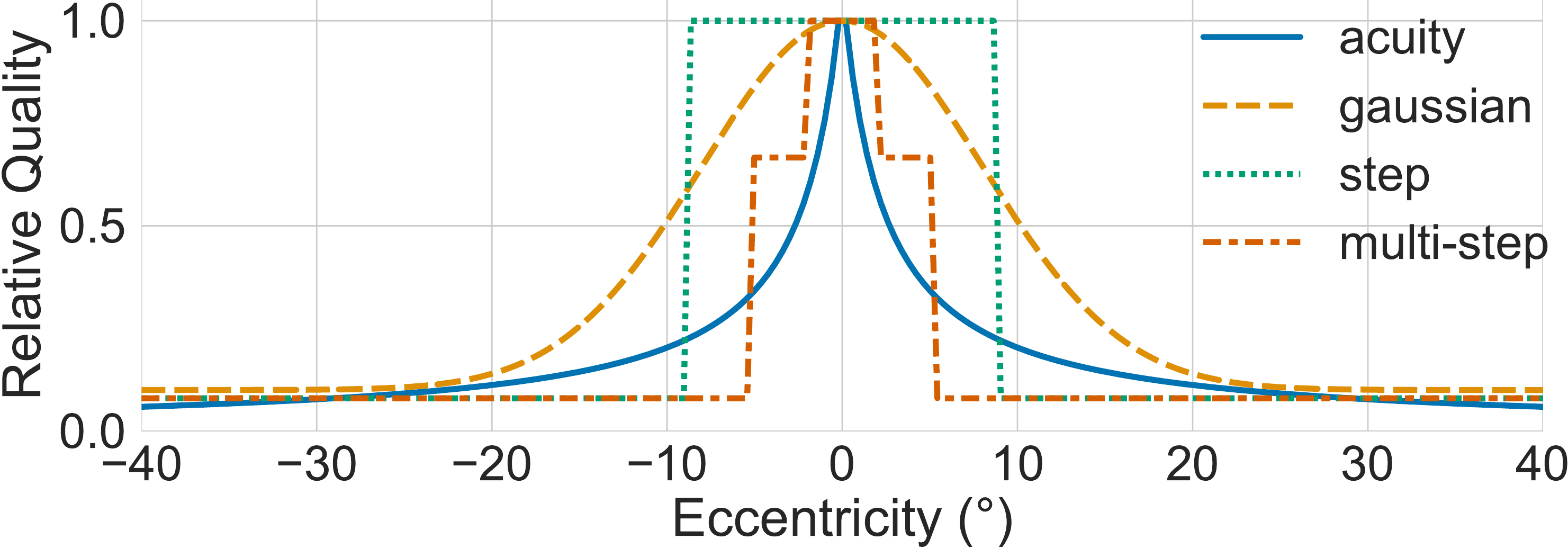}
    \caption{
        Related work uses a variety of functions to approximate relative quality (i.e., the allocation of bits) with the decay in visual acuity.
    }\label{fig:shapes}
    \vspace{-1.5em}
\end{figure}

Foveated video compression relies on accurate, real-time gaze information to allocate a larger portion of the bitrate to where a viewer is looking while decaying the quality in the periphery.
Assuming accurate and instantaneous gaze information, these algorithms can compress frames to have minimally-sized regions of high-resolution without viewer detection.
In practice, however, latency introduces uncertainty in a viewer's gaze position, requiring larger regions of high-resolution video\footnote{Inaccuracy in an eye tracking device also contributes to this uncertainty but is out of scope of this work.}.
\Cref{fig:approximations} illustrates this challenge.
On the left, the gaze position used by the system matches the actual gaze position perfectly, and the periphery can be highly compressed.
However, a system must also keep the foveal region large enough such that when the gaze moves, it does not escape the region before the system can react (shown on the right).
This occurs if the system latency, $t_L$, is longer than the time it takes for the gaze to move.
Consequently, there is tension between minimally sizing the foveal region for better compression and sizing it large enough to ensure a viewer does not see video artifacts.

While we understand the decay in visual acuity of the human visual system well~\cite{geisler1998real, robson1981probability}, our understanding of the nuances of peripheral vision (e.g., change blindness, crowding, object recognition, etc.) is still actively developing~\cite{strasburger2011peripheral, rosenholtz2016capabilities}.
Because of these nuances, there is no well-understood mapping function that a foveated compression algorithm can use to transmit the minimal number of bits while maintaining high visual quality for all types of videos.

As a result, foveation is usually achieved by empirically choosing an approximation function to model the decay in visual acuity and applying transformations that appear visually acceptable.
For example, Illahi~et~al.~\cite{illahi2020cloud} and Wiedemann~et~al.~\cite{wiedemann2020foveated} choose a Gaussian function, Romero~et~al.~\cite{romero2018foveated} choose a step function, and Guenter~et~al.~\cite{guenter_foveated_2012} choose a step function with multiple steps.
\Cref{fig:shapes} shows examples of these approximations functions.

\Cref{fig:teaser,fig:shapes} also plot the acuity model of Geisler~et~al.~\cite{geisler1998real}, fit with parameters from Robson~et~al.~\cite{robson1981probability}, in blue.
This gives visual acuity, $A$, as a function of eccentricity, $e$, as follows.
\begin{equation}
    A(e) = \ln(64)\frac{2.3}{0.106 * (e + 2.3)}
\end{equation}

The goal of these approximations is to minimize the gap between the transmitted quality and the perceived quality.
As annotated in \Cref{fig:teaser}, transmitting too high of a quality in the periphery wastes bits while transmitting too low of a quality results in visual artifacts.
Further, as system latency increases, so does uncertainty about the viewer's gaze and, consequently, the size of the foveal region.
The approximation functions must be widened to accommodate this uncertainty, resulting in more wasted bandwidth.

\begin{figure*}[t]
    \centering
    \includegraphics[width=0.95\textwidth]{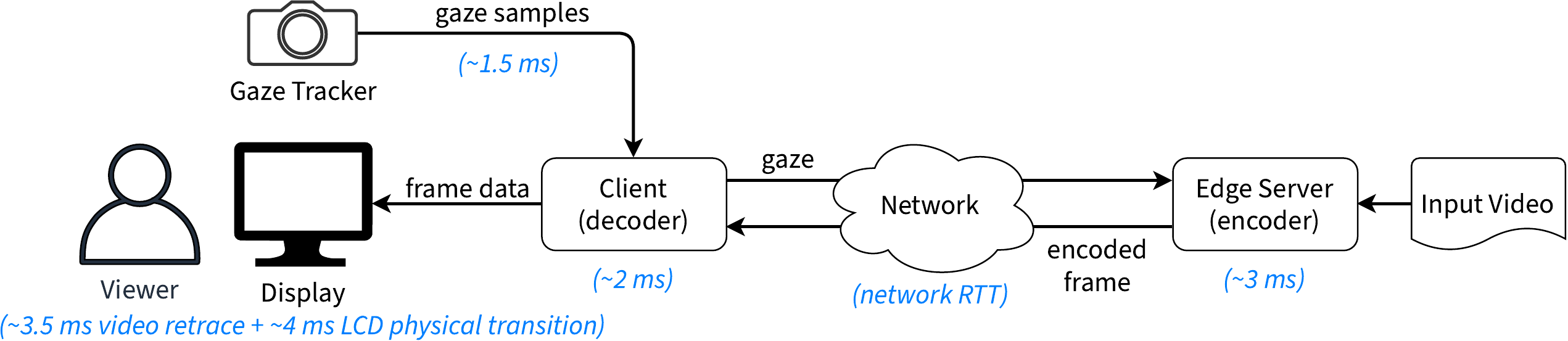}
    \caption{%
        Overview of our low-latency, desktop-based prototype system.
        This system allows us to focus on the effects of latency on foveated video compression by avoiding the limitations complexities of current VR HMDs.
    }\label{fig:overview}
    \vspace{-1em}
\end{figure*}

In practice, we find that the choice of approximation function also influences implementation choices, which can in and of itself cause additional latency.
For example, a common foveated compression implementation of a Gaussian approximation is to vary the degree of compression of individual subregions of a video frame according to the Gaussian function~\cite{illahi2020cloud, wiedemann2020foveated}.
This requires processing the full video resolution to produce a single video stream of smoothly varying quality.
In contrast, a simple step function can be implemented using two traditionally compressed video streams---one for the cropped high-resolution foveal region and one for the low resolution background.
This approach results in far less processing.
For example, rather than processing a full 4K (\SI{3840 x 2160}{px}) video, a two-stream approach might process a small \SI{480 x 480}{px} foveal region and a downscaled \SI{768 x 432}{px} background, which combines to be \SI{< 7}{\percent} of the original 4K pixels.

To focus on the impact of reducing latency, we chose a simple two-stream approach (\Cref{sec:system}).
Our experience suggests that achieving low latencies will be key to realistically achieving retina-quality VR video over a network.
We cannot have long latencies and achieve great compression; we need great latencies as well.

\section{A Low-Latency Prototype System}%
\label{sec:system}

Understanding the real-world impact of latency on foveated video compression requires a system with very low latencies.
However, commercial head-mounted displays (HMDs) used for VR today have system latencies \SI{>45}{\ms}~\cite{stein2021comparison}.
In addition, these HMDs do not have sufficiently high resolutions (i.e., less than 4K) to be an ideal test bed for studying the impact of latency on compression of retina-quality video\footnote{There are upcoming HMDs, such as the Vive Pro 2, which will include 4k or higher resolution displays.}.
Consequently, we build a desktop-based system as a proxy for future VR HMDs.
Doing so allows us to focus on the impact of latency without the limitations of current HMDs.

\subsection{Architecture}
We design our system based on a typical video-streaming architecture with a client and server model.
However, rather than the client only receiving encoded video frames from the server to decode and display, the client also sends the viewer's current gaze position each time a frame is received (\Cref{fig:overview}).
This gaze sample allows the server to encode the next video frame foveated on the viewer's gaze position.
To minimize system latency, the server and client run as separate processes on the same machine and communicate using message passing, implemented with shared memory.

\subsection{Two-Stream Compression}
To reduce the latency spent on encoding and decoding, our system uses a simple two-stream approach.
The server sequentially reads uncompressed frames at the frame rate of the input video.
Then, for each gaze sample it receives from the client, it compresses up to two versions of the current frame\footnote{We also skip both background frames when the current video frame has not changed and foreground frames if the gaze has not not changed.}.
First, it downscales the video frame to a significantly lower resolution.
Second, it crops the video frame to a small area around the viewer's gaze location.
The resolution of both the downscale and the crop are configurable.
It then encodes these two frames to send to the client.
At the client, the reverse process occurs.
First, the client decodes and upscales the background frame to the size of its display.
Next, it decodes the foreground frame and positions it at the corresponding gaze position with a blend\footnote{We set the alpha channel (opacity) to a 2D Gaussian in order to fade out the hard, square edges of the foreground. The parameters of the Gaussian are chosen empirically.}.
Finally, it displays this composed frame.

As is typical with compression techniques, this approach trades off increased computation (real-time encoding per client) for reduced bitrate.
While the server can pre-encode the background, the foreground must be encoded in real-time using the viewer's gaze.

\subsection{System Details}%
\label{sec:system_details}

We implement our system in Rust, using SDL2, FFmpeg, and x264.
Our workstation runs Pop!\_OS 20.04 and contains an AMD Ryzen 7 3700X CPU, \SI{16}{\giga\byte} of memory, and an NVIDIA GeForce RTX 2070 SUPER GPU.
Our display is an LG 27GN95B-B (4K at \SI{144}{\hertz}, \SI{7.6}{\ms} input latency).
An Eyelink 1000 provides low-latency eye tracking.
The software for this system available at \url{https://github.com/lukehsiao/fvideo}.

\section{Experiments}

Using our low-latency prototype system as a proxy for future VR HMDs, we seek to answer the following questions.
\begin{enumerate}
    \item What is the lower bound for latency of modern hardware?
    \item What is the latency of our foveated compression system, and where is the time spent?
    \item What is the relationship between system latency and achievable video compression?
\end{enumerate}

\subsection{Lower Bound for System Latency}

\begin{figure}[t]
    \centering
    \setlength{\fboxsep}{0pt}
    \fbox{\includegraphics[width=0.8\columnwidth]{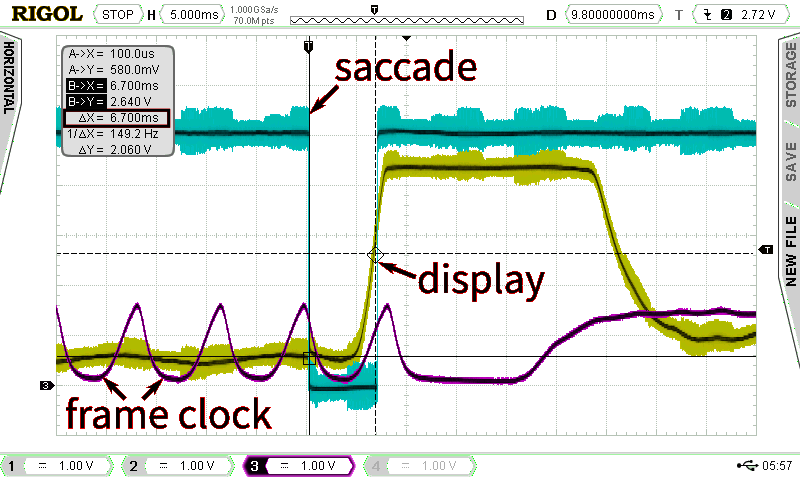}}
    \caption{%
        We use an oscilloscope to measure a lower bound for system latency---the time between an artificial saccade occurring (the falling edge in green) and the pixels of the display reacting (the rising edge in yellow).
    }\label{fig:zisworks_latency}
    \vspace{-1.0em}
\end{figure}

\begin{figure}[t]
    \centering
    \includegraphics[width=1.0\columnwidth]{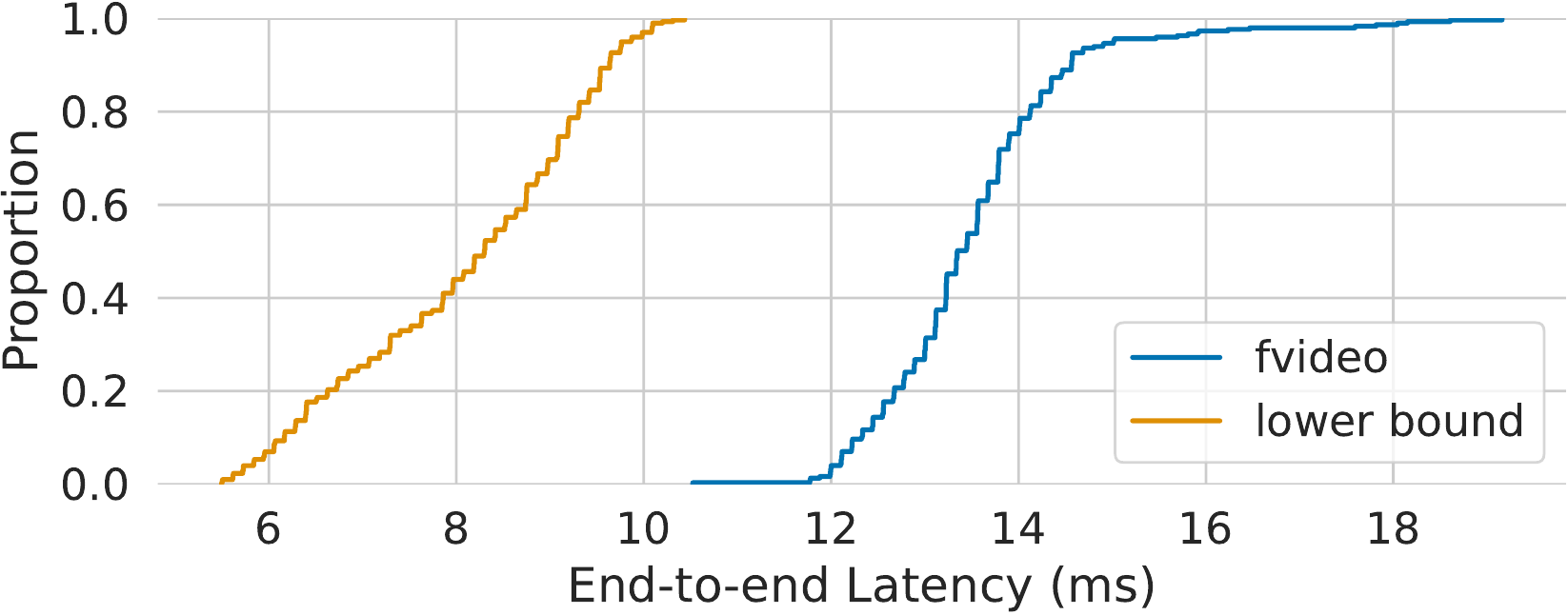}
    \caption{%
        ECDF of end-to-end system latencies.
        A simple two-stream approach for compression (fvideo) only adds \SI{\sim 5}{\ms} over the lower bound.
    }\label{fig:compression_cdf}
    \vspace{-1.5em}
\end{figure}

The first experiment finds a lower bound for the achievable system latency using commercially available hardware.
We use an eye tracker and display that are among the lowest latency available today and minimize video processing by only toggling portions of the display between black and white (i.e., omitting video encoding/decoding).
We use an Eyelink 1000 to minimize the latency between a viewer's eyes moving and receiving the data in software.
Although lower-latency eye trackers are continually being developed~\cite{Angelopoulos2021Gaze}, the Eyelink 1000 provides a good trade-off between accuracy, latency\footnote{We disable the built-in filters to further minimize latency}, and sampling rate among those that are commercially available\footnote{Based on their advertised specifications and prior comparison by others~\cite{stein2021comparison}.}.

To minimize the latency between a frame being sent to the display and the pixels changing, we select a ZisWorks x28 R2 monitor (1080p resolution at \SI{240}{\hertz}), which advertises an input latency of \SI{\sim 30}{\micro\second}, significantly lower than the \SIrange{1.5}{16}{\ms} of most consumer monitors.
We also opt for a simple graphics stack for this experiment by using Xubuntu 18.04 with compositing disabled.

To ensure measurements are precise, automated, and repeatable, we design our own Arduino-based artificial saccade generator (ASG)\footnote{See \url{https://github.com/lukehsiao/eyelink-latency}. Unable to find a suitable commercial ASG, we follow the precedent of related work by building our own.}.
Most eye trackers (head-mounted or desktop) either track the infrared (IR) reflection of the retina or directly process a video stream of the eye to detect and track the pupil~\cite{reingold2014eye}.
The Eyelink 1000 uses IR reflection, so we build an ASG that can be triggered using software and toggles between two IR LEDs to mimic a saccade.

Finally, we implement a minimal system that polls for changes in gaze position using the eye tracker and then uses OpenGL to change a small portion of the display from black to white.
This pixel change is then detected using a photodiode circuit.
The approach of using an ASG and photodiode circuit to measure latency is commonly used~\cite{bernard2007page, bockisch1999different, reingold2014eye}.
System latency is measured as the time between triggering the ASG and the mid-point of seeing the pixel change on the photodiode.
\Cref{fig:zisworks_latency} shows an oscilloscope trace of this process with a system latency of \SI{6.7}{\ms}.
In some cases, it is possible for the saccade to be triggered and the display pixels to change within the one refresh cycle of the monitor.
However, if the saccade does not line up with the frame clock, then it may take up to an additional refresh cycle to update.

We run this measurement for 300 repetitions and plot the empirical cumulative distribution function (ECDF) in \Cref{fig:compression_cdf} (lower bound).
The minimum observed latency is under \SI{6}{\ms}, with the majority of samples falling under \SI{9}{ms}.
Of this latency, an average of \SI{1.65}{\ms} is waiting for the updated gaze sample, and the remaining is dominated by the time it takes for the display to update (\SI{>4}{\ms}).

\subsection{Foveated Compression Latency}

Next, we measure the latency of our gaze-contingent, foveated compression prototype.
There are two important differences in this experiment compared to the previous lower bound baseline.
First, this experiment includes the computational cost of scaling, cropping, encoding, and decoding 4K video frames.
Rather than directly changing a portion of a frame from black to white, we use a synthetic video.
This video is black until a saccade is detected, after which it toggles a portion of the frame to white.
Second, this experiment uses the LG 27GN95B-B monitor, which supports 4K resolution at \SI{144}{\hertz} and has an average of \SI{7.6}{\ms} input latency.

We also run this measurement 300 times and plot the ECDF in \Cref{fig:compression_cdf} (fvideo).
On average, our system has \SI{\sim{}5}{\ms} longer latency than our lower bound baseline.
Of this additional latency, \SI{\sim{}2.8}{\ms} comes from the slower display, and the remaining comes from the computational costs of encoding and decoding the two video streams.
Importantly, our two-stream approach ensures that the latency of foveated video compression itself is not significantly longer than the latency of our hardware.
An approximate breakdown of where time is spent is annotated in \Cref{fig:overview}.
A gaze sample is taken and sent to the edge server, where a new frame is encoded.
This frame is then sent to the client for decoding and display.

\subsection{User Study: Latency vs.\ Bitrate}%
\label{sec:user_study}

We conduct a user study to better understand the relationship between system latency and how much compression can be achieved while maintaining similar visual quality.
We set up a controlled laboratory experiment to gather data on perceived video quality using our low-latency prototype (\Cref{sec:system}).

Because this work is motivated by the challenge of streaming retina-quality VR video (\Cref{sec:intro}), we use a 4K video encoded at \SI{28}{\mega\bit\per\second} as a proxy for the video quality of streaming platforms like YouTube\footnote{Specifically, we use \texttt{x264 {-}{-}preset veryfast {-}{-}bitrate 28000}.}.
In this study, we measure the compressed bitrate of a video as a function of system latency at the point of equipoise perceived video quality compared to the baseline.

As stimuli, we use two 4K videos from Derf's collection~\cite{derfcollection}.
These two videos are selected due to their diverse content, and each consists of two sub-scenes.
The first, \texttt{barscene}, shows one sub-scene with strong bokeh and another with dialogue between two individuals that naturally guides a viewer's gaze.
The second, \texttt{square\_timelapse}, shows one sub-scene of a busy crowd of people where viewers' gaze typically jumps around the scene, and another of a city skyline with many hard edges and natural scenery.
We test these videos at three latency conditions.
First, we evaluate our system at its unmodified latency (\SI{\sim 14}{\ms}).
Then, we select the minimum and maximum latencies of commercially available HMDs as measured by Stein~et~al.~\cite{stein2021comparison}: \SIlist{45;81}{\ms}.
We add artificial delay to our system to match these latencies.

For each video and latency combination, we prepare a set of compression configurations starting at lower resolutions with higher compression, and moving to higher resolutions with lower compression.
We only evaluate \num{3} latency points and \num{2} videos, to keep the study to a reasonable duration.

\subsubsection{Experimental Setup}
We use the system detailed in \Cref{sec:system_details}.
The LG display is set to \SI{3840 x 2160}{px} and \SI{144}{\hertz}.
Physically, the display is \SI{59.67 x 33.56}{\cm} and placed at a distance that gives \ang{\sim 55} horizontal field of view, achieving retina-quality resolution.
The participant's head is stabilized using a chin and forehead rest, and the eye tracker is placed between the monitor and participant.


\subsubsection{Procedure}
Each participant was asked to view two videos and perform the same task on each.
The order in which the videos were presented was equally divided among participants.
We first calibrated the eye tracker and validated the tracking accuracy for each participant.
Then, participants were asked to perform four trials of a matching task.
Three of the trials correspond to each latency points (\SIlist{14;45;81}{\ms}), and we randomly repeated one trial to check for consistency.
The order of the trials was also randomized.
Participants were shown a reference video and then asked to select which of ten comparison videos the reference video is most similar to in quality for each trial.
The ten videos were ordered by increasing video quality.
Participants could also choose to respond that none of the ten were similar in quality.
Participants were allowed to take as long as they wished to make their selection and could freely navigate and re-watch any of the videos.
Each participant did \SI{8}{trials}, resulting in a study duration of \SI{\sim 45}{\minute}.

\subsubsection{Participants}
We recruited 13 participants\footnote{The COVID-19 pandemic limited the number of participants available for this study.}.
All participants provided written consent before taking part in the study, and the methods were approved by Stanford's institutional review board (IRB).
Before each experiment, the participants were briefed about the purpose of the study and their task.
Of these 13 participants, we excluded 2 participants' data from the results because we were unable to achieve a maximum calibration accuracy error \ang{< 10} (unacceptably large compared to the size of the foveal region).

\subsubsection{Results}
\begin{figure}[t]
    \centering
    \includegraphics[width=1.0\columnwidth]{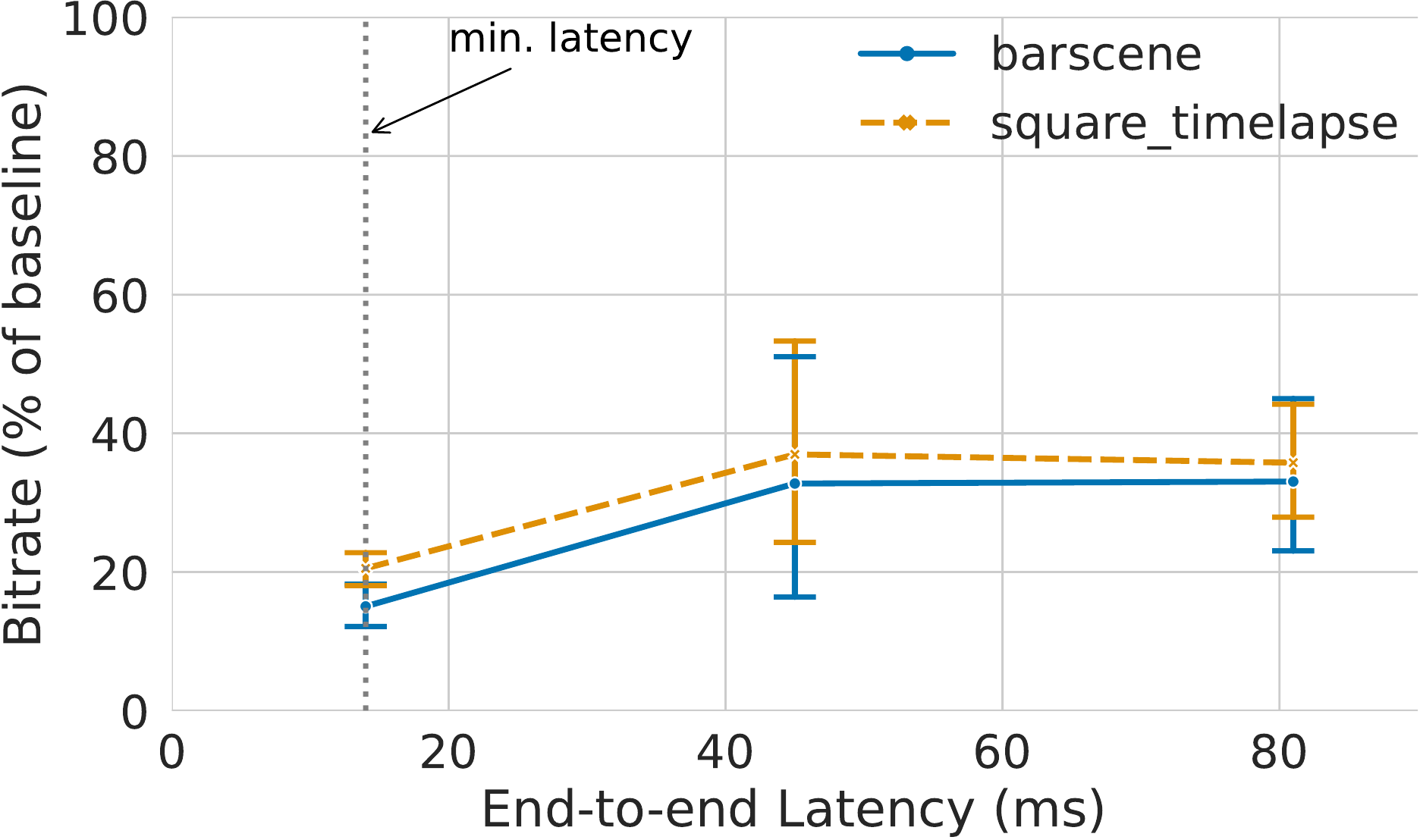}
    \caption{%
        Latency vs.\ compression, with \SI{95}{\percent}-confidence intervals.
        We improve compression by \SI{5}{\times} using a simple two-stream approach.
        However, the full benefit comes only at latencies lower than demonstrated by current VR HMDs.
    }\label{fig:user_study}
    \vspace{-1.0em}
\end{figure}

\Cref{fig:user_study} shows the results of our user study.
We plot the mean compressed bitrate as a percentage of the \SI{28}{\mega\bit\per\second} baseline along with the \SI{95}{\percent}-confidence interval for each latency.
At low latency, a simple two-stream approach can compress these videos to \SI{\sim{}20}{\percent} of the baseline while maintaining a similar visual quality.
Despite using a simple algorithm, our results are competitive with the numbers reported in related work (\Cref{sec:foveated_compression}).

To understand the statistical significance of these differences, we compute a t-test between both the \SIlist{14;45}{\ms} latencies and \SIlist{45;81}{\ms} latencies.
We find that the difference between the means of \SIlist{14;45}{\ms} is statistically significant ($t=2.76$, $p=0.008$), while the difference between the means of \SIlist{45;81}{\ms} is not ($t=0.10$, $p=0.92$).
This validates the trend shown in \Cref{fig:user_study}.

The latency gap between the fastest and slowest consumer HMDs is a significant \SI{36}{\ms}.
However, we find that reducing the latency from \SI{81}{\ms} to \SI{45}{\ms} does not significantly improve the required video bitrate.
It is not until we push the system's latency to below that of commercially available HMDs that we see an additional \SI{\sim 2}{\times} compression benefit.
This finding also suggests this relationship is not simply a question of making the foveal region larger as the delay increases---we suspect there is a distinct phenomenon (and a compression opportunity) at low latencies.

Related works that mention system latency often do so primarily to show that the latency is below the \SI{\sim 50}{\ms} proposed by prior work (\Cref{sec:background}).
However, our finding not only suggests that driving down system latency can result in significant compression gains without changing the compression algorithm itself, but also that these gains might only be realized with system latencies much lower than previously proposed thresholds.

\section{Conclusion}%
\label{label:conclusion}

We present latency reduction as a method for improving foveated video compression and validate its potential by implementing a prototype, ultra-low-latency video streaming system.
Our findings indicate that reducing system latency is greatly helpful to achieving the levels of compression needed for retina-quality VR content.
Although the techniques presented here cannot enable retina-quality VR video alone, they serve as a data point and an early step towards that goal.
The impact of latency must be studied in more environments (e.g., on VR HMDs and over real networks), using more applications (e.g., a slew of foveated graphics techniques), and with more content (e.g., a more diverse set of videos).

The latency budget we describe is tight, but in a model where an edge server can be located within a few milliseconds RTT of the client, we believe server-side video rendering at retina quality may become feasible at practical network throughputs.
In concert with future advancements in VR HMDs and improvements in foveated video compression, reducing latency may play a critical role in making retina-quality VR video streaming practical over realistic communication networks.

\begin{acks}
This work was supported by Facebook Reality Labs, by NSF grants 2045714, 1909212, 2039070, and 1839974, by Google, VMware, Dropbox, and Amazon, and by a Stanford Knight-Hennessy Fellowship, an Okawa Research Grant and a Sloan Research Fellowship.
\end{acks}

\bibliographystyle{ACM-Reference-Format}
\bibliography{references}

\end{document}